\documentclass[article,shortnames,nojss]{jss}\usepackage[]{graphicx}\usepackage[]{color}
\makeatletter
\def\maxwidth{ %
  \ifdim\Gin@nat@width>\linewidth
    \linewidth
  \else
    \Gin@nat@width
  \fi
}
\makeatother

\definecolor{fgcolor}{rgb}{0.345, 0.345, 0.345}
\newcommand{\hlnum}[1]{\textcolor[rgb]{0.686,0.059,0.569}{#1}}%
\newcommand{\hlstr}[1]{\textcolor[rgb]{0.192,0.494,0.8}{#1}}%
\newcommand{\hlopt}[1]{\textcolor[rgb]{0,0,0}{#1}}%
\newcommand{\hlstd}[1]{\textcolor[rgb]{0.345,0.345,0.345}{#1}}%
\newcommand{\hlkwb}[1]{\textcolor[rgb]{0.69,0.353,0.396}{#1}}%
\newcommand{\hlkwc}[1]{\textcolor[rgb]{0.333,0.667,0.333}{#1}}%
\newcommand{\hlkwd}[1]{\textcolor[rgb]{0.737,0.353,0.396}{\textbf{#1}}}%

\usepackage{framed}
\makeatletter
\newenvironment{kframe}{%
 \def\at@end@of@kframe{}%
 \ifinner\ifhmode%
  \def\at@end@of@kframe{\end{minipage}}%
  \begin{minipage}{\columnwidth}%
 \fi\fi%
 \def\FrameCommand##1{\hskip\@totalleftmargin \hskip-\fboxsep
 \colorbox{shadecolor}{##1}\hskip-\fboxsep
     \hskip-\linewidth \hskip-\@totalleftmargin \hskip\columnwidth}%
 \MakeFramed {\advance\hsize-\width
   \@totalleftmargin\z@ \linewidth\hsize
   \@setminipage}}%
 {\par\unskip\endMakeFramed%
 \at@end@of@kframe}
\makeatother

\definecolor{shadecolor}{rgb}{.97, .97, .97}
\definecolor{messagecolor}{rgb}{0, 0, 0}
\definecolor{warningcolor}{rgb}{1, 0, 1}
\definecolor{errorcolor}{rgb}{1, 0, 0}
\newenvironment{knitrout}{}{} % an empty environment to be redefined in TeX

\usepackage{alltt}
\usepackage{amssymb}
\usepackage{amsmath}
\usepackage{natbib}
\usepackage{graphicx}
\usepackage{color}
\usepackage{verbatim}
\usepackage{hyperref}
\usepackage{url}
\usepackage{multicol,lipsum}
\usepackage{multirow}
\usepackage{nameref}

\author{Craig Wang\\
Department of Mathematics\\
University of Zurich, Switzerland \And
Reinhard Furrer\\
Department of Mathematics\\
Department of Computational Science\\
University of Zurich, Switzerland}

\title{\pkg{eggCounts}: a Bayesian hierarchical toolkit to model faecal egg count reductions}

\Plainauthor{Craig Wang, Reinhard Furrer} %% comma-separated
\Plaintitle{eggCounts: a Bayesian hierarchical approach to model faecal egg count reduction} %% without formatting
\Shorttitle{\pkg{eggCounts}: a Bayesian hierarchical toolkit to model FECR} %% a short title (if necessary)

\Abstract{
This is a vignette for the \proglang{R} package \pkg{eggCounts} version 2.0. The package implements a suite of Bayesian hierarchical models dealing with faecal egg count reductions (FECR). The models are designed for a variety of practical situations, including individual treatment efficacy, zero inflation, small sample size (less than 10) and potential outliers. The functions are intuitive to use and their outputs are easy to interpret, such that users are protected from being exposed to complex Bayesian hierarchical modelling tasks. In addition, the package includes plotting functions to display data and results in a visually appealing manner. The models have been implemented in \proglang{Stan} modelling language, which provides efficient sampling technique to obtain posterior samples. This vignette briefly introduces different models and provides a short walk-through analysis with example data.
}
\Plainkeywords{Bayesian hierarchical model, treatment efficacy, anthelmintic resistance, R package rstan}

\Address{
  Craig Wang\\
  Department of Mathematics\\
  University of Zurich, Zurich, Switzerland\\
  E-mail: \email{craigwang247@gmail.com}\\ \\

  Reinhard Furrer\\
  Department of Mathematics, Department of Computational Science\\
  University of Zurich, Zurich, Switzerland\\
  E-mail: \email{reinhard.furrer@uzh.ch} \\
  URL: \url{http://user.math.uzh.ch/furrer/} \\
}
\IfFileExists{upquote.sty}{\usepackage{upquote}}{}
\begin{document}
%\SweaveOpts{concordance=TRUE}
%\SweaveOpts{prefix.string=vignette}
%%%%%%%%%%%%%%
%%%% Section 1
%%%%%%%%%%%%%%

\section{Introduction}

The prevalence of anthelmintic resistance in livestock has increased in recent years, as a result of the extensive use of anthelmintic treatments to reduce infection of parasitic worms. Parasite infection can pose a large economic burden on ruminant production if it is left uncontrolled \citep{Perry_1999}, hence it is crucial to monitor treatment efficacy via accurate and reliable methods. The faecal egg count reduction test (FECRT) is commonly applied to estimate reduction and its confidence interval, it was suggested in the World Association for the Advancement of Veterinary Parasitology (WAAVP) guideline \citep{Coles1992}. The test computes reduction using the ratio of after- and before-treatment means, and calculates its confidence interval using the asymptotic variance of their log ratio. Recently, several authors have shown that the FECRT is not capable to address some practical problems. Counting techniques such as McMaster \citep{Coles1992} uses a low analytical sensitivity, this introduces substantial variability in results which are not accounted for by the FECRT \citep{Torgerson2012}. As a consequence, the estimated efficacy or percentage of egg count reduction was found to be quiet variable, especially in samples with low before-treatment faecal egg counts (FECs). Other high-sensitivity counting techniques such as FLOTAC \citep{FLOTAC2010} and Cornell-Wisconsin \citep{Egwang1982} can reduce but not completely eliminate the variability \citep{Torgerson2012, Levecke2012194}. Further, the distribution of eggs tends to be aggregated within the host population \citep{Grenfell1995}. \cite{Levecke2012} pointed out results from FECRT should be interpreted with caution when aggregation level is high. \cite{Pena2016} showed the coverage probability is suboptimal in high-aggregation settings. Finally, the FECRT cannot be used to compute a confidence interval of reductions when all of the after-treatment counts are zero.

Over recent years, there is an emerging trend of using Bayesian hierarchical models to analyze FECR \citep{NEVES2014, GEURDEN2015, KRUCKEN2017, Pyziel2018}. Those analysis are typically done via either online user-friendly graphical interface (original: \cite{Torgerson2014}, updated: \cite{shinyegg}) or dedicated \proglang{R} packages. To the best of our knowledge, there are only two existing packages on \proglang{CRAN}, namely \pkg{eggCounts} \citep{egg} and \pkg{bayescount} \citep{bayescount} that analyze FECR. One of the key differences between those two packages is their underlying assumptions. \pkg{eggCounts} assumes analytical sensitivity-adjusted gamma-Poisson distributions, where the gamma distribution captures aggregation of FECs between animals and the Poisson distribution captures sampling variation. \pkg{bayescount} assumes compound gamma-gamma-Poisson distributions, where the sampling variation is represented by the gamma-Poisson (or negative binomial) distribution. Both packages provide standard models for common scenarios such as paired and unpaired setting, zero-inflation and individual efficacy, however, they also have some non-overlapping functionalities. In particular, \pkg{eggCounts} provides additional models for 1) small sample size and 2) counts with potential outliers, while \pkg{bayescount} provides additional models for 1) varying aggregation level and 2) repeated counts; and tools for power analysis. \cite{Wang2018} compared their model performance under different scenarios in a simulation study.

The Bayesian hierarchical models in \pkg{eggCounts} are implemented with \proglang{Stan} modelling language via \pkg{rstan} package \citep{rstan}, which uses No-U-Turn Sampler (NUTS), an improved version of Hamiltonian Monte Carlo \citep{Homan2014} to efficiently obtain posterior samples of model parameters. It is computationally advantageous and has easy-to-interpret syntaxes. Using Bayesian hierarchical models can be challenging for non-statistical specialists \citep{Matthews}. This vignette aims to bridge the gap between the need to use reliable statistical methods to evaluate FECR and the amount of resources available to guide using such methods. The remainder of this vignette is organized as follows. Section \ref{sec:load} provides information about how to install and load the software. Section \ref{sec:models} introduces model formulations. Section \ref{sec:data} provides a data analysis example on FECR using example data. Finally, Section \ref{sec:diss} concludes with a short discussion.

%%%%%%%%%%%%%%
%%%% Section 2
%%%%%%%%%%%%%%
\section{Loading and using the software}
\label{sec:load}
\pkg{eggCounts} is a package for the statistical software \proglang{R} \citep{R}. The software is available for Linux, Windows and macOS operating systems, and can be freely downloaded from the Comprehensive R Archive Network (CRAN, \url{https://cloud.r-project.org/}).

Both \pkg{eggCounts} and \pkg{bayescount} rely on external tools that need to be installed separately. While the latter relies on \proglang{JAGS}, \pkg{eggCounts} relies on \proglang{Stan}. Hence it is necessary for users to install \proglang{Stan} on their operating system. Detailed installation instructions can be found via the official \pkg{rstan} wiki via

\begin{center}
\url{https://github.com/stan-dev/rstan/wiki/RStan-Getting-Started}
\end{center}

Once ready, \pkg{eggCounts} can be installed and loaded using the commands

\begin{knitrout}
\definecolor{shadecolor}{rgb}{0.969, 0.969, 0.969}\color{fgcolor}\begin{kframe}
\begin{alltt}
\hlkwd{install.packages}\hlstd{(}\hlstr{"eggCounts"}\hlstd{)}
\hlkwd{library}\hlstd{(eggCounts)}
\end{alltt}
\end{kframe}
\end{knitrout}

To check if functions are working properly, and to see a working pipeline for evaluating FECR with \pkg{eggCounts}, a short demo can be run with the command

\begin{knitrout}
\definecolor{shadecolor}{rgb}{0.969, 0.969, 0.969}\color{fgcolor}\begin{kframe}
\begin{alltt}
\hlkwd{demo}\hlstd{(fecm_stan,} \hlkwc{package}\hlstd{=}\hlstr{"eggCounts"}\hlstd{)}
\end{alltt}
\end{kframe}
\end{knitrout}

%%%%%%%%%%%%%%
%%%% Section 3
%%%%%%%%%%%%%%
\section{Modelling faecal egg count reduction}
\label{sec:models}
This section outlines different Bayesian hierarchical models implemented in \pkg{eggCounts} for analyzing faecal egg count data. Two non-Bayesian approaches that are implemented in \pkg{eggCounts} are also mentioned. The majority of the models are implemented for evaluating egg count reductions, the primary output contains summary statistics regarding the estimated reduction.

\subsection{Preliminary}
We define some terms and notations that will be used throughout this vignette.
\subsubsection{Unpaired design}
Suppose there are two groups of animals: a control group with sample size $n_C$ which did not receive anthelmintic treatment, and a treatment group with sample size $n_T$. A faecal sample from each animal is collected and counted. This is the \textit{unpaired} design.
\subsubsection{Paired design}
Suppose there is a group of animals with sample size $n$. A faecal sample from each animal within the group is collected once before treatment and once some days after treatment. This is the \textit{paired} design.
\subsubsection{Notation}
Table \ref{tab:not} contains notations and their definitions used in the baseline models. We use index $i$ to denote the $i$th sample or the $i$th animal. Additional model-specific notations will be introduced in the subsequent model descriptions.
\begin{table}[h]
\centering
\begin{tabular}{cll}
\hline
                  & \textbf{Notation}        & \textbf{Definition} \\ \hline
\multirow{5}{*}{Sample level} & $Y_i^{*C}$ & Observed counts before treatment (control group)\\
                             & $Y_i^{*T}$ & Observed counts after treatment (treatment group)\\
                             & $Y_i^C$    & True epg from collected before treatment sample \\
                             & $Y_i^T$    & True epg from collected after treatment sample \\ \
                             & $f_i$          & Analytical sensitivity \\ \hline
\multirow{1}{*}{Individual level} & $\mu_i$    & Individual latent mean epg   \\ \hline
\multirow{3}{*}{Group level} & $\delta$      & Proportion of epg remaining (treatment efficacy) \\
                             & $\mu$         & Group latent mean epg   \\
                             & $\kappa$      & Dispersion of faecal eggs between animals \\      \hline
\end{tabular}
\caption{Table of notations for the two baseline models.}
\label{tab:not}
\end{table}

\subsection{Two baseline models}
The two simplest models are shown in Equation~\eqref{eq:base1} and Equation~\eqref{eq:base2}. They are the building blocks of more advanced models that we introduce later. The models can be divided into three layers:
\begin{enumerate}
\item the Binomial distributions capture counting variability;
\item the Poisson distributions address Poisson error, which arises because of randomly distributed eggs within the faecal sample; and
\item the Gamma distribution captures FEC aggregation between animals.
\end{enumerate}
The baseline models assume the same reduction $\delta$ is experienced by each animal within the group.

\begin{multicols}{2}
\center \textbf{Unpaired design}
\begin{equation}
\begin{aligned}
\label{eq:base1}
Y^{*C}_i|Y^C_i &\sim \text{Binomial}(Y^C_i,1/f_i),\\
Y^{*T}_{i}|Y^T_i &\sim \text{Binomial}(Y^T_i,1/f_i),\\
Y^C_i|\mu^C_i &\sim \text{Poisson}(\mu^C_i),\\
Y^T_i|\mu^T_i &\sim \text{Poisson}(\delta\mu^T_i),\\
\mu^C_i|\kappa,\mu &\sim \text{Gamma}(\kappa,\kappa/\mu).\\
\mu^T_i|\kappa,\mu &\sim \text{Gamma}(\kappa,\kappa/\mu).
\end{aligned}
\end{equation}

\columnbreak

\center \textbf{Paired design}
\begin{equation}
\begin{aligned}
\label{eq:base2}
Y^{*C}_i|Y^C_i &\sim \text{Binomial}(Y^C_i,1/f_i),\\
Y^{*T}_{i}|Y^T_i &\sim \text{Binomial}(Y^T_i,1/f_i),\\
Y^C_i|\mu^C_i &\sim \text{Poisson}(\mu^C_i),\\
Y^T_i|\mu^C_i &\sim \text{Poisson}(\delta\mu^C_i),\\
\mu^C_i|\kappa,\mu &\sim \text{Gamma}(\kappa,\kappa/\mu).\\
\end{aligned}
\end{equation}
\end{multicols}

\begin{figure}[h]
\center

\includegraphics[width=\textwidth]{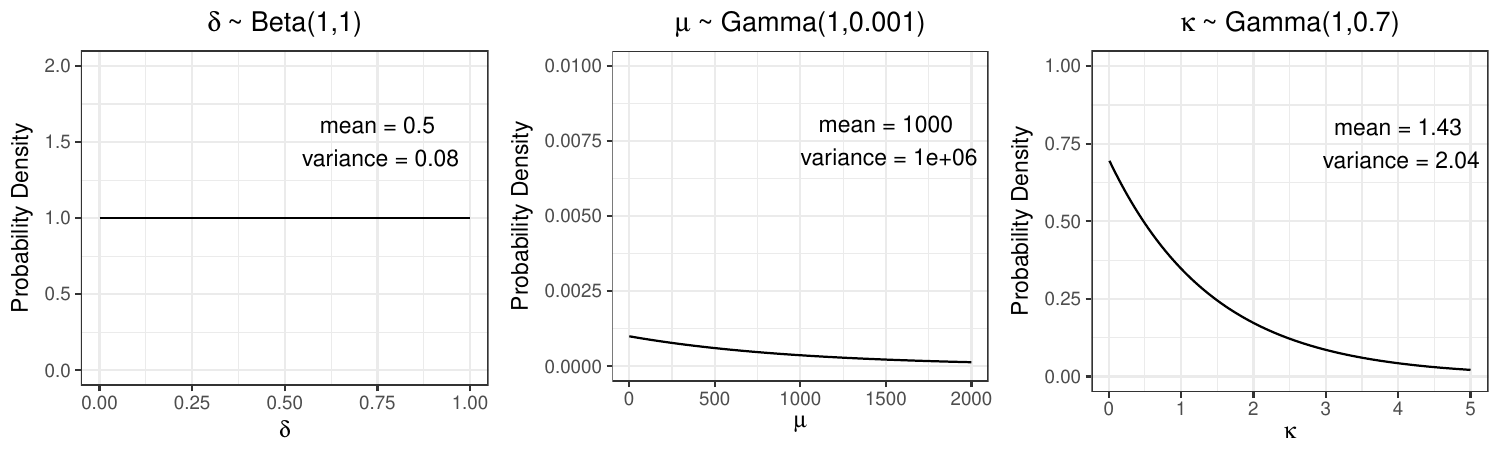}
\caption{Default priors for the baseline models.}
\label{fig:priors}
\end{figure}

Since the models are in Bayesian framework, priors are required for the parameters $\delta, \mu$ and $\kappa$. The priors shown in Figure~\ref{fig:priors} are used by default. Users can also supply their own priors in a list format, for example, setting the argument \code{muPrior = list(priorDist = "normal", hyperpars=c(1000,100))} in \code{fecr_stan()} assigns a $\textrm{Normal}(1000,100^2)$ prior to $\mu$.

\subsection{Model extensions}
\subsubsection{Individual treatment effect}
\code{fecr_stan(..., paired = TRUE, zeroInflation = FALSE, indEfficacy = TRUE)}\\
\cite{Pena2016} and \cite{levecke2018} pointed out the limitation of the baseline model for paired design, namely assuming the same reduction $\delta$ for each animal. As a result of this limitation, the baseline model was extended by \cite{Wang2018} to allow each animal having different treatment efficacies $\delta_i$. The model explicitly uses the paired relationships to estimate reductions, and it can effectively model before- and after-treatment aggregation level changes as well.

The modified parts of the baseline model is shown in Equation~\eqref{eq:ind}. The efficacies $\delta_i$ follow a gamma distribution with shape $\tau$ and rate $\tau/\nu$.
\begin{center}
\textbf{Paired design}
\end{center}
\begin{equation}
\begin{aligned}
\label{eq:ind}
Y^T_i|\mu^C_i &\sim \text{Poisson}(\delta_i\mu^C_i),\\
\delta_i|\tau,\nu &\sim \text{Gamma}(\tau,\tau/\nu).\\
\end{aligned}
\end{equation}
A Beta(1,1) prior is assigned to $\nu$, and a zero-truncated Normal(2,1) prior is assigned to $\tau$. The group median reduction is used as the reduction estimate, which has been shown to perform well in a comprehensive simulation study \citep{Wang2018}. For identifiability reasons, this extension only applies to the paired design.

\subsubsection{Zero inflation}
\code{fecr_stan(..., zeroInflation = TRUE, indEfficacy = FALSE)}\\
\cite{Wang2017} introduced the model variation to allow zero-inflated true mean epg, which can arise from a mixture of infected and unexposed animals. Instead of Poisson distributed true epg from collected samples, they follow zero-inflated Poisson distribution with zero-inflation parameter $\phi$. Using this model for data without underlying zero-inflated distribution does not have a negative impact on the performance. The modified parts of the baseline model are shown in Equation~\eqref{eq:zero1} and Equation~\eqref{eq:zero2}.

\begin{multicols}{2}
\center \textbf{Unpaired design}
\begin{equation}
\begin{aligned}
\label{eq:zero1}
Y^C_i|\mu^C_i &\sim \text{ZIPoisson}(\mu^C_i, \phi),\\
Y^T_i|\mu^T_i &\sim \text{ZIPoisson}(\delta\mu^T_i, \phi),\\
\end{aligned}
\end{equation}

\columnbreak

\center \textbf{Paired design}
\begin{equation}
\begin{aligned}
\label{eq:zero2}
Y^C_i|\mu^C_i &\sim \text{ZIPoisson}(\mu^C_i, \phi),\\
Y^T_i|\mu^C_i &\sim \text{ZIPoisson}(\delta\mu^C_i, \phi),\\
\end{aligned}
\end{equation}
\end{multicols}

A Beta(1,1) prior is assigned to $\phi$.

\subsubsection{Small sample size}
\label{chap:sss}
\underline{Informative priors}
\code{fecr_stan(..., muPrior = , deltaPrior = ,...)}\\
When sample size is less than 10, an automatic warning message is prompted to the user at end of model output. \pkg{eggCounts} has functions \code{getPrior_mu()} and \code{getPrior_delta()} to help users determining the prior parameters for $\mu$ and $\delta$ based on some quantitative belief.

For both $\mu \sim \text{Gamma}(\theta_1, \theta_2)$ and $\delta \sim \text{Beta}(\theta_1, \theta_2)$, the prior parameters can be found via quantile matching estimation by solving $(\theta_1, \theta_2)$ in,
\begin{equation}
\begin{aligned}
F^{-1}(p_1 | \theta_1, \theta_2) &= Q_1, \\
F^{-1}(p_2 | \theta_1, \theta_2) &= Q_2,
\end{aligned}
\end{equation}
where $F^{-1}$ is an inverse cumulative distribution function from either a gamma or a beta distribution, $p_1$ and $p_2$ are probabilities at corresponding quantiles $Q_1$ and $Q_2$. In addition for $\delta \sim \text{Beta}(\theta_1, \theta_2)$, \code{getPrior_delta()} can obtain its prior parameters from its mode and concentration by
\begin{equation}
\begin{aligned}
\theta_1 &= \omega \cdot (k-2) + 1, \\
\theta_2 &= (1 - \omega) \cdot (k-2) + 1,
\end{aligned}
\end{equation}
where $\omega$ is the mode and $k$ is the concentration parameter of a beta distribution.

\underline{Simplified model}
\code{fecr_stanSimple(...)}\\
In the context of very small samples a simpler model with less parameters could be beneficial. Small samples contribute very limited information to the estimation of dispersion parameter~$\kappa$, dropping the parameter removes the gamma-layer and reduces complexity of the model. From a practical point of view, this means that there are no aggregation of FECs between animals, or at least not observable with a small number of animals. The modified parts of the baseline model are shown in Equation~\eqref{eq:simple}.
\begin{center}
\textbf{Paired design}
\end{center}
\begin{equation}
\begin{aligned}
\label{eq:simple}
Y^C_i|\mu &\sim \text{Poisson}(\mu),\\
Y^T_i|\mu &\sim \text{Poisson}(\delta\mu),\\
\end{aligned}
\end{equation}

\subsubsection{One-sample model}
\code{fec_stan(...)}\\
Instead of modeling egg count reductions, the one-sample model can be used when the interest is merely estimating the number of egg counts. The same gamma-Poisson distribution (or its zero-inflated version) is used but now only with a single set of observed counts $Y^{*}_i$.
\begin{equation}
\begin{aligned}
Y^{*}_i|Y_i &\sim \text{Binomial}(Y_i,1/f_i),\\
Y_i|\mu_i &\sim \text{Poisson}(\mu_i),\\
\mu_i|\kappa,\mu &\sim \text{Gamma}(\kappa,\kappa/\mu).\\
\end{aligned}
\end{equation}

\subsubsection{Data with outliers}
\label{chap:dwo}
\code{fecr_stanExtra(..., modelName = c("Po", "UPo", "ZIPo", "ZIUPo"), ...)}\\
Additional models are available externally for handling FECs with potential outliers or bi-modality, that is, having counts that are clearly separated from the ``normal'' population. The models are in \pkg{eggCountsExtra} package hosted on Github. The stan model codes can be loaded for modelling using the command,
\begin{knitrout}
\definecolor{shadecolor}{rgb}{0.969, 0.969, 0.969}\color{fgcolor}\begin{kframe}
\begin{alltt}
\hlstd{devtools}\hlopt{::}\hlkwd{install_github}\hlstd{(}\hlstr{"CraigWangStat/eggCountsExtra"}\hlstd{)}
\hlkwd{library}\hlstd{(eggCountsExtra)}
\end{alltt}
\end{kframe}
\end{knitrout}
then apply \code{fecr_stanExtra()} function from \pkg{eggCounts}. There are two outliers and weight definitions.
\begin{itemize}
\item Unpaired design: Compute the mean of after-treatment counts excluding those higher than $Q3 + 1.5 \cdot IQR$, where Q3 is the 75th percentile and IQR is the inter-quartile range. After-treatment counts that are higher than 95th percentile of Poisson distribution with the computed mean are classified as outliers. Non-outliers are assigned with weight 1, while the highest outlier is assigned with weight 0.01 and other outliers follow proportionally.
\item Paired design: Animals with an increased after-treatment counts are classified as outliers. Non-outliers are assigned with weight 1, while outliers are assigned with weight equal to the ratio of before- and after-treatment count.
\end{itemize}
The modified parts of the baseline model are shown in Equation~\eqref{eq:out1} and Equation~\eqref{eq:out2}.

\begin{multicols}{2}
\center \textbf{Unpaired design}
\begin{equation}
\begin{aligned}
\label{eq:out1}
Y^T_i|\mu^T_i \sim & \; w_i \cdot \text{Poisson}(\delta\mu^T_i) + \\
& (1-w_i)\cdot\text{Poisson}(\alpha \cdot \delta\mu^T_i),\\
\end{aligned}
\end{equation}

\columnbreak

\center \textbf{Paired design}
\begin{equation}
\begin{aligned}
\label{eq:out2}
Y^T_i|\mu^C_i \sim & \; w_i \cdot \text{Poisson}(\delta\mu^C_i) + \\
& (1-w_i)\cdot\text{Poisson}(\alpha \cdot \delta\mu^C_i),\\
\end{aligned}
\end{equation}
\end{multicols}

where $w_i$ are the weights and $\alpha$ is the scaling factor for outliers. An additional weighted Poisson component is also added in the zero-inflated cases for handling outliers. A one-truncated Normal$(\bar{y}_o^{*T}/\bar{y}^{*T}, 10^2)$ prior is assigned to $\alpha$, where $\bar{y}_o^{*T}$ is the weighted mean of outliers and $\bar{y}^{*T}$ is the mean of all after-treatment counts.

\subsubsection{Custom models}
\code{fecr_stanExtra(..., modelCode = , ...)}\\
\code{fecr_stanExtra(..., modelFile = , ...)}\\
If advanced users are desired to supply their own models and use the functions that are already in \pkg{eggCounts} package, \code{fecr_stanExtra()} can be used to run the analysis. One of \code{modelCode} and \code{modelFile} argument need to be supplied for this purpose. The code template is available in \pkg{eggCountExtra} package and it can be inspected by the command,
\begin{knitrout}
\definecolor{shadecolor}{rgb}{0.969, 0.969, 0.969}\color{fgcolor}\begin{kframe}
\begin{alltt}
\hlstd{devtools}\hlopt{::}\hlkwd{install_github}\hlstd{(}\hlstr{"CraigWangStat/eggCountsExtra"}\hlstd{)}
\hlkwd{library}\hlstd{(eggCountsExtra)}
\hlkwd{writeLines}\hlstd{(}\hlkwd{getTemplate}\hlstd{())}
\end{alltt}
\end{kframe}
\end{knitrout}
The model provided need to be consistent with the parameter naming conventions in the template.

\subsection{Inference and diagnostics}
All models in \pkg{eggCounts} package are fitted within Bayesian framework using MCMC simulation. The models are implemented in \proglang{Stan} modelling language via the \pkg{rstan} package, which are based on compiled \proglang{C++} code. By executing wrapper functions \code{fec_stan}, \code{fecr_stan} or \code{fecr_stanExtra}, sampling algorithm are launched in the background and model results are printed via the \proglang{R} Console. Arguments for Markov chains can be supplied via those wrapper functions, including \code{nsamples, nburnin, thinning, nchain, ncore} and \code{adaptDelta}. Generally, the default values for those arguments are applicable for most problems.

Model diagnostics are automatically conducted on the posterior samples to ensure the results are reliable. Undesirable behaviors of the Markov chains may occur when the data is difficult to model.  The joint posterior distribution is not sufficiently explored when there are divergent transitions after warmup, a warning message will be printed in the console when this occurs. The tuning parameter \code{adaptDelta} should be increased to mitigate this problem. The convergence of Markov chains is checked via the potential scale reduction factors \citep{Brooks1998}, a warning message will also be printed if there is evidence for non-convergence. While the printed information in the console is sufficient for most users, we offer the possibility to examine a more detailed model output by setting \code{saveALL = TRUE} and extract \code{stan.samples} from the model output list.

\subsection{Non-Bayesian approaches}
\subsubsection{FECRT}
\code{fecrtCI(...)} \\
The FECRT \citep{Coles1992} is implemented according to the WAAVP guideline.
\begin{equation}
\label{eq:fecrt}
\text{Percentage reduction } = 100\times \left(1-\frac{\bar{y}_T}{\bar{y}_C}\right),
\end{equation}
where $\bar{y}_T$ and $\bar{y}_C$ denote the mean counts of the treatment and the control group. Assuming independence, the estimated asymptotic variance of the log ratio is given by
\begin{equation}
\label{fecrt_ci}
 \widehat{\text{Var}}\left(\log{\frac{\bar{Y}_T}{\bar{Y}_C}}\right)=\frac{s^2_T}{n_T \bar{y}^2_T}+\frac{s^2_C}{n_C \bar{y}^2_C}.
\end{equation}
where $\bar{Y}_T$ and $\bar{Y}_C$ denote the means of random samples, $s^2_T$ and $s^2_C$ denote the sample variances. The variance can be used to construct an approximate 95\% CI of the log ratio using the 2.5\% and the 97.5\% quantile of a Student's t-distribution with $n_T+n_C-2$ degrees of freedom. The 95\% CI for the estimated reduction can be obtained via transformation.

\subsubsection{Non-parametric bootstrap}
\code{fecrtCI(...)} \\
Each bootstrap sample is generated by resampling the data with replacement. The reduction of each sample is evaluated using Equation~\eqref{eq:fecrt}. The estimated reduction and its confidence interval are then computed based on the estimated reductions of all bootstrap samples.

%%%%%%%%%%%%%%
%%%% Section 4
%%%%%%%%%%%%%%

\section{Example data analysis}
\label{sec:data}
In this example, we run the individual efficacy model for the paired design without zero inflation on an example dataset.

\begin{knitrout}
\definecolor{shadecolor}{rgb}{0.969, 0.969, 0.969}\color{fgcolor}\begin{kframe}
\begin{alltt}
\hlkwd{set.seed}\hlstd{(}\hlnum{1}\hlstd{)}
\hlstd{simdf} \hlkwb{<-} \hlkwd{simData2s}\hlstd{(}\hlkwc{n} \hlstd{=} \hlnum{15}\hlstd{,} \hlkwc{preMean} \hlstd{=} \hlnum{500}\hlstd{,} \hlkwc{delta} \hlstd{=} \hlnum{0.1}\hlstd{,} \hlkwc{kappa} \hlstd{=} \hlnum{1}\hlstd{,}
                   \hlkwc{f} \hlstd{=} \hlnum{15}\hlstd{,} \hlkwc{paired} \hlstd{=} \hlnum{TRUE}\hlstd{)}
\hlkwd{head}\hlstd{(simdf,} \hlnum{3}\hlstd{)}
\end{alltt}
\begin{verbatim}
##   obsPre masterPre truePre obsPost masterPost truePost
## 1     75         5      66       0          0        6
## 2   1050        70     954     150         10      100
## 3    915        61     943      60          4       81
\end{verbatim}
\end{kframe}
\end{knitrout}

\begin{knitrout}
\definecolor{shadecolor}{rgb}{0.969, 0.969, 0.969}\color{fgcolor}\begin{kframe}
\begin{alltt}
\hlkwd{plotCounts}\hlstd{(simdf[,}\hlkwd{c}\hlstd{(}\hlstr{"obsPre"}\hlstd{,}\hlstr{"obsPost"}\hlstd{)])}
\end{alltt}
\end{kframe}
\end{knitrout}
\begin{figure}[ht]
\center
\includegraphics[width=0.5\textwidth]{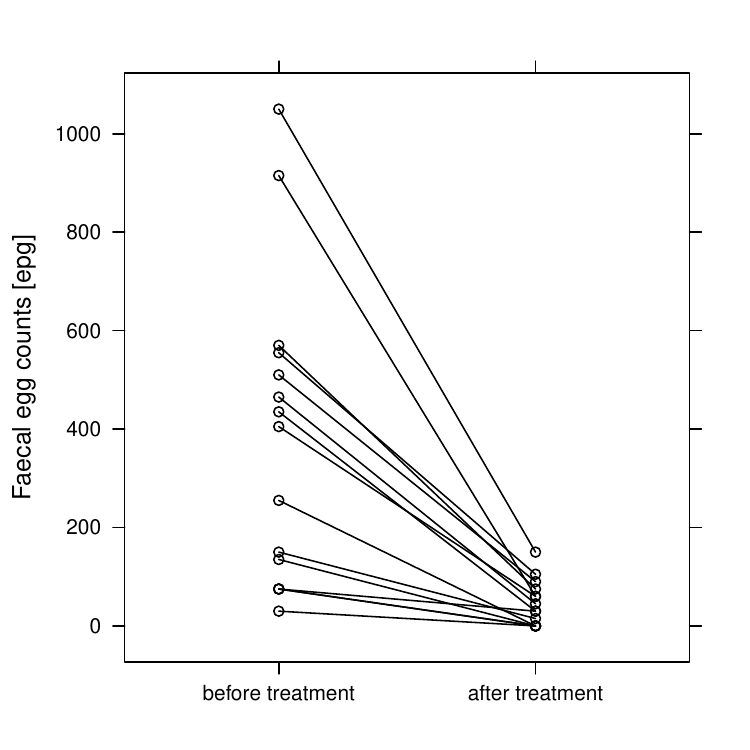}
\caption{Pairwise before and after treatment epg.}
\label{fig:eg1data}
\end{figure}

The simulated dataset consists of 15 paired samples, with a true epg of 500 before treatment and a true reduction of 90\%. The \code{truePre} and \code{truePost} columns indicate the true epg in the obtained before- and after-treatment samples. Figure~\ref{fig:eg1data} indeed shows a large reduction for all observed samples. For estimating the FECR, we can either use the \code{masterPre} and \code{masterPost} column with argument \code{rawCounts = TRUE} or use the \code{obsPre} and \code{obsPost} column with argument \code{rawCounts = FALSE}.

\begin{knitrout}
\definecolor{shadecolor}{rgb}{0.969, 0.969, 0.969}\color{fgcolor}\begin{kframe}
\begin{alltt}
\hlstd{model} \hlkwb{<-} \hlkwd{fecr_stan}\hlstd{(simdf}\hlopt{$}\hlstd{obsPre, simdf}\hlopt{$}\hlstd{obsPost,} \hlkwc{rawCounts} \hlstd{=} \hlnum{FALSE}\hlstd{,} \hlkwc{preCF} \hlstd{=} \hlnum{15}\hlstd{,}
                   \hlkwc{paired} \hlstd{=} \hlnum{TRUE}\hlstd{,} \hlkwc{zeroInflation} \hlstd{=} \hlnum{FALSE}\hlstd{,} \hlkwc{indEfficacy} \hlstd{=} \hlnum{TRUE}\hlstd{)}
\end{alltt}
\end{kframe}
\end{knitrout}

\begin{CodeOutput}


Model:  Bayesian model without zero-inflation for paired design allowing
        individual efficacy
 Number of Samples:  2000
 Warm-up samples:  1000
 Thinning:  1
 Number of Chains 2
                      mean       sd     2.5%      50%
FECR                0.8955   0.0268   0.8341   0.8981
meanEPG.untreated 422.5811 115.1893 254.1534 404.2357
meanEPG.treated    44.1221  16.8003  20.8828  41.3992
                     97.5% HPDLow95     mode HPDHigh95
FECR                0.9406   0.8439   0.9006    0.9455
meanEPG.untreated 693.0552 226.5460 375.3411  654.0410
meanEPG.treated    86.0950  19.2576  39.0260   79.7146

NOTE: There are no evidence of non-convergence since all parameters have
      potential scale reduction factors less than 1.1.
\end{CodeOutput}

There are no warning messages from the model output and there is no evidence of non-convergence. Hence we can report the output: there is an 89.8\% reduction with a 95\% equal-tailed credible interval of (83.4, 94.1). Next, we can compute the probability that the reduction is less than some threshold, say 95\%, based on the posterior density of the reduction.

\begin{knitrout}
\definecolor{shadecolor}{rgb}{0.969, 0.969, 0.969}\color{fgcolor}\begin{kframe}
\begin{alltt}
\hlkwd{fecr_probs}\hlstd{(model}\hlopt{$}\hlstd{stan.samples,} \hlkwc{plot} \hlstd{=} \hlnum{FALSE}\hlstd{)}
\end{alltt}
\begin{verbatim}
## The probability that the reduction is less than 0.95 is 99.25 %.
\end{verbatim}
\end{kframe}
\end{knitrout}

\begin{figure}[ht]
\center
\includegraphics[width=0.85\textwidth]{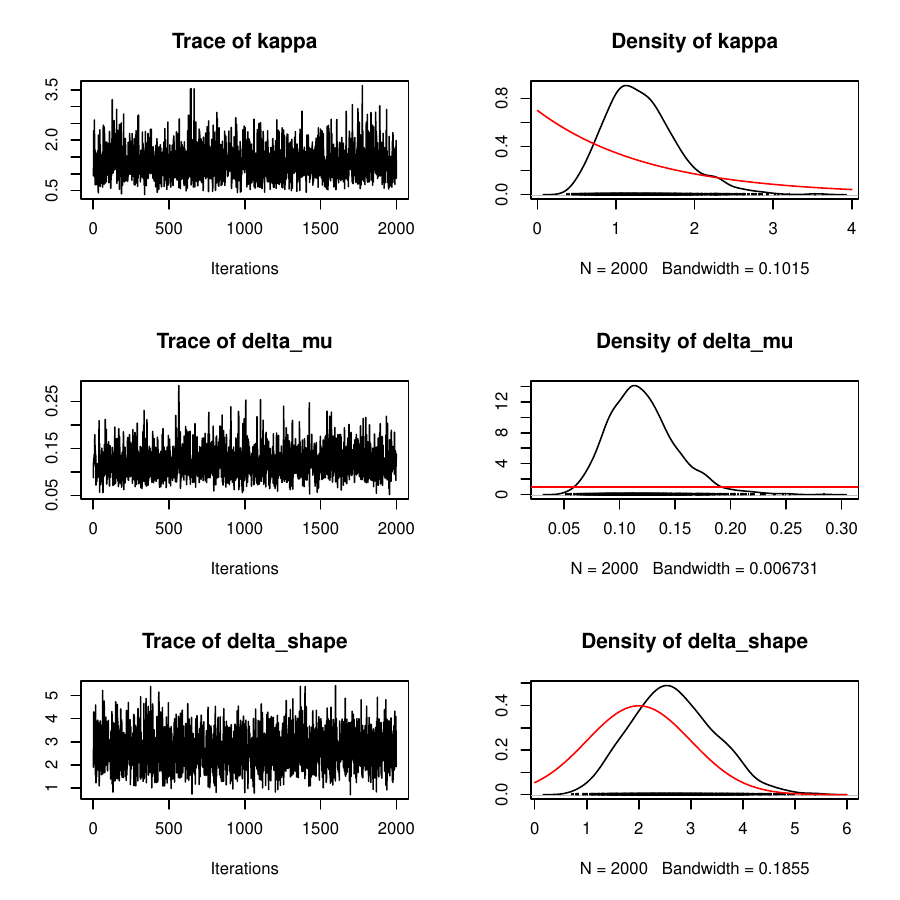}
\caption{Traceplots and posterior densities of selected parameters from the paired model with individual efficacy. The priors are shown in red lines in the density plots.}
\label{fig:eg1trace}
\end{figure}

In the case of any doubt, the posterior samples of relevant parameters can be extracted and investigated further. For example, we can apply the function \code{stan2mcmc()} to obtain a \code{mcmc} object and use \pkg{coda} package to take a look at the traceplots and densities with the code below. The outputs are shown in Figure~\ref{fig:eg1trace}.

\begin{knitrout}
\definecolor{shadecolor}{rgb}{0.969, 0.969, 0.969}\color{fgcolor}\begin{kframe}
\begin{alltt}
\hlstd{samples} \hlkwb{<-} \hlkwd{stan2mcmc}\hlstd{(model}\hlopt{$}\hlstd{stan.samples)}
\hlkwd{par}\hlstd{(}\hlkwc{mfcol}\hlstd{=}\hlkwd{c}\hlstd{(}\hlnum{3}\hlstd{,}\hlnum{2}\hlstd{))}
\hlkwd{plot}\hlstd{(samples[,}\hlkwd{c}\hlstd{(}\hlstr{"kappa"}\hlstd{,}\hlstr{"delta_mu"}\hlstd{,}\hlstr{"delta_shape"}\hlstd{)],} \hlkwc{density}\hlstd{=}\hlnum{FALSE}\hlstd{,}
     \hlkwc{auto.layout}\hlstd{=}\hlnum{FALSE}\hlstd{)}
\hlkwd{plot}\hlstd{(samples[,}\hlstr{"kappa"}\hlstd{],} \hlkwc{trace}\hlstd{=}\hlnum{FALSE}\hlstd{,} \hlkwc{auto.layout}\hlstd{=}\hlnum{FALSE}\hlstd{,}
     \hlkwc{main}\hlstd{=}\hlstr{"Density of kappa"}\hlstd{)}
\hlkwd{lines}\hlstd{(x}\hlkwb{<-}\hlkwd{seq}\hlstd{(}\hlnum{0}\hlstd{,}\hlnum{4}\hlstd{,}\hlnum{0.01}\hlstd{),} \hlkwd{dgamma}\hlstd{(x,} \hlnum{1}\hlstd{,} \hlnum{0.7}\hlstd{),} \hlkwc{col}\hlstd{=}\hlstr{"red"}\hlstd{)}
\hlkwd{plot}\hlstd{(samples[,}\hlkwd{c}\hlstd{(}\hlstr{"delta_mu"}\hlstd{)],} \hlkwc{trace}\hlstd{=}\hlnum{FALSE}\hlstd{,} \hlkwc{auto.layout}\hlstd{=}\hlnum{FALSE}\hlstd{,}
     \hlkwc{main}\hlstd{=}\hlstr{"Density of delta_mu"}\hlstd{)}
\hlkwd{lines}\hlstd{(x}\hlkwb{<-}\hlkwd{seq}\hlstd{(}\hlnum{0}\hlstd{,}\hlnum{1}\hlstd{,}\hlnum{0.01}\hlstd{),} \hlkwd{dbeta}\hlstd{(x,} \hlnum{1}\hlstd{,} \hlnum{1}\hlstd{),} \hlkwc{col}\hlstd{=}\hlstr{"red"}\hlstd{)}
\hlkwd{plot}\hlstd{(samples[,}\hlkwd{c}\hlstd{(}\hlstr{"delta_shape"}\hlstd{)],} \hlkwc{trace}\hlstd{=}\hlnum{FALSE}\hlstd{,} \hlkwc{auto.layout}\hlstd{=}\hlnum{FALSE}\hlstd{,}
     \hlkwc{main}\hlstd{=}\hlstr{"Density of delta_shape"}\hlstd{)}
\hlkwd{lines}\hlstd{(x}\hlkwb{<-}\hlkwd{seq}\hlstd{(}\hlnum{0}\hlstd{,}\hlnum{6}\hlstd{,}\hlnum{0.01}\hlstd{),} \hlkwd{dnorm}\hlstd{(x,} \hlnum{2}\hlstd{,} \hlnum{1}\hlstd{),} \hlkwc{col}\hlstd{=}\hlstr{"red"}\hlstd{)}
\end{alltt}
\end{kframe}
\end{knitrout}

\section{Discussion}
\label{sec:diss}
This vignette has introduced the \proglang{R} package \pkg{eggCounts}, which can fit a number of Bayesian hierarchical models that are designed to estimate FECR in different scenarios, including individual treatment efficacy, zero-inflation, small sample size and potential outliers. By utilizing the \proglang{Stan} modelling language, the models are computationally faster than conventional sampling algorithms. The functions are tailored to users without extensive statistical training. The streamlined model outputs are straightforward to interpret, and automatic model diagnostic procedures are implemented to report any concerns.

It is important to be aware of the assumptions corresponding to each model, in order to obtain reliable results. For instance, the baseline model assumes the same efficacy for each animal. A strong violation of this assumption will lead to the underestimated variance of the posterior distribution for $\delta$. It is also recommended to check the appropriateness of the priors against the data at hand. For example, the Beta(1,1) prior limits the reduction to be between 0\% and 100\%. If an increase in epg is observed in many animals, a uniform prior with an upper bound higher than 1 should be assigned to the reduction parameter.

We kindly ask users to provide feedback on the \pkg{eggCounts} package. If there are any concerns about the model validity or interpretations of model output, please seek statistical advice to ensure reliable results are obtained.

\section*{Acknowledgements}
We thank Anja Fallegger and Tea Isler for their contributions to the \hyperref[chap:sss]{Small sample size} and \hyperref[chap:dwo]{Data with outliers} models during their master theses, and thank Roman Flury for his comments on this vignette. \\ \\
Last updated: 03/02/2022

\bibliography{eggvig}
\end{document}